\documentclass[10pt,twoside]{article}

\usepackage[dvipdfmx]{graphicx,color}
\usepackage[dvipsnames]{xcolor}
\usepackage[font={small},labelsep=space]{caption}
\usepackage{graphbox,subcaption}
\usepackage{amsmath,amssymb, mathrsfs,bm,bold-extra}
\usepackage{booktabs,multirow,dcolumn,bigdelim}
\usepackage{fancybox,arydshln,mdframed,setspace}
\usepackage{enumitem,sectsty, titlesec}
\usepackage[hang,flushmargin]{footmisc}
\usepackage{pdflscape,afterpage,capt-of}
\usepackage{endnotes,filecontents,etoolbox}
\usepackage{here,stackengine,empheq,cases,braket}
\usepackage[TU]{fontenc}
\usepackage{fancyhdr}
\usepackage[makeroom]{cancel}
\usepackage{mwe,ulem}
\usepackage{newtxtext,newtxmath}
\usepackage{comment,endnotes}

\usepackage[compress,authoryear,round]{natbib}
\setcitestyle{notesep={;},aysep={,},yysep={;}}

\usepackage[font={footnotesize,stretch=1.2},labelsep=space]{caption}
\usepackage{subcaption}

\DeclareCaptionLabelFormat{vert}{\textbf{{#1} {#2}~|}}
\newcommand{\f}[2]{\frac{#1}{#2}}

\newcommand{\abs}[1]{\left| #1 \right|}
\newcommand{\bmt}[1]{{{\mbox{\boldmath$ #1 $}}}}
\newcommand{\q}[1]{`#1'}

\renewcommand*{\thepage}{\footnotesize\arabic{page}}
\newcolumntype{d}{D{.}{.}{2.4}}

\newcommand{\rn}[1]{\uppercase\expandafter{\romannumeral #1\relax}}

\def\X{\mathrm{X}}
\def\Y{\mathrm{Y}}
\def\cK{\mathcal{K}}

\newcommand{\disc}[2]{\href{https://explore.openalex.org/concepts/#1}{\textcolor{BlueViolet}{#2}}}

\def\pol{\disc{c17744445}{Political science}}
\def\phil{\disc{c138885662}{Philosophy}}
\def\econ{\disc{162324750}{Economics}}
\def\biz{\disc{c144133560}{Business}}
\def\psy{\disc{c15744967}{Psychology}}
\def\math{\disc{c33923547}{Mathematics}}
\def\med{\disc{c71924100}{Medicine}}
\def\bio{\disc{c86803240}{Biology}}
\def\comp{\disc{c41008148}{Computer science}}
\def\geol{\disc{c127313418}{Geology}}
\def\chem{\disc{c185592680}{Chemistry}}
\def\art{\disc{c142362112}{Art}}
\def\soc{\disc{c144024400}{Sociology}}
\def\engi{\disc{c127413603}{Engineering}}
\def\geog{\disc{c205649164}{Geography}}
\def\hist{\disc{c95457728}{History}}
\def\mate{\disc{c192562407}{Materials science}}
\def\phys{\disc{c121332964}{Physics}}
\def\envi{\disc{c39432304}{Environmental science}}

\usepackage[hyphens]{xurl}
\usepackage[pdfencoding=auto,linkbordercolor=CornflowerBlue,citecolor=tb,linkcolor=tb,pdfpagelabels=false]{hyperref}
\hypersetup{
  colorlinks=true,
  allbordercolors=CornflowerBlue
}

\def\mtitle{Evolving landscape of US--China science collaboration: Convergence and divergence}

\newif\ifabbreviation
\pretocmd{\thebibliography}{\abbreviationfalse}{}{}
\AtBeginDocument{\abbreviationtrue}
\DeclareRobustCommand\acroauthor[2]{%
  \ifabbreviation #2\else #1\fi}

\def\bibfont{\small}
\makeatletter
\newcommand*{\myfnsymbol}[1]{\ensuremath{%
\ifcase#1 \or \ast \or \dagger \or \ddagger \or \spadesuit \or \diamondsuit \or \clubsuit \or \heartsuit \else \@ctrerr \fi}}
\makeatother

\definecolor{tb}{rgb}{0.24, 0.43, 0.91}
\renewcommand*{\cite}[1]{\textcolor{tb}{\citep{#1}}}
\newcommand*{\citek}[1]{\textcolor{tb}{\citet{#1}}}

\begin{document}

\quad\vspace{-1.4cm}
\begin{flushright}
September 2023 [v1]
\end{flushright}

\vspace{0.5cm}

\begin{center}
\fontsize{13pt}{14pt}\selectfont\bfseries
\mtitle
\end{center}

\renewcommand*{\thefootnote}{\textcolor{black}{$\myfnsymbol{\value{footnote}}$}}

\vspace*{0.8cm}
\centerline{%
{Kensei Kitajima}\,\footnote{\,\tt kitajima@nistep.go.jp | \href{https://orcid.org/0000-0001-8799-4332}{orcid.org/0000-0001-8799-4332}}${}^{;\,1}$
~~and~~
{Keisuke Okamura}\,\footnote{\,\tt okamura@ifi.u-tokyo.ac.jp | \href{https://orcid.org/0000-0002-0988-6392}{orcid.org/0000-0002-0988-6392}}${}^{;\,2,\,3}$
}

\vspace*{0.6cm}
{\small
\centerline{\textit{%
${}^{1}$National Institute of Science and Technology Policy (NISTEP),}}
\centerline{\textit{%
3-2-2 Kasumigaseki, Chiyoda-ku, Tokyo 100-0013, Japan.}}
\vspace*{3mm}
\centerline{\textit{%
${}^{2}$Institute for Future Initiatives (IFI), The University of Tokyo,}}
\centerline{\textit{%
7-3-1 Hongo, Bunkyo-ku, Tokyo 113-0033, Japan.}}
\vspace*{3mm}
\centerline{\textit{%
${}^{3}$SciREX Center, National Graduate Institute for Policy Studies (GRIPS),}}
\centerline{\textit{%
7-22-1 Roppongi, Minato-ku, Tokyo 106-8677, Japan.}}
\vspace*{0.5cm}
}

\vspace{1.0cm}
\noindent\textbf{Abstract.}
\quad
International research collaboration among global scientific powerhouses has exhibited a discernible trend towards convergence in recent decades. 
Notably, the US and China have significantly fortified their collaboration across diverse scientific disciplines, solidifying their status as a national-level duopoly in global scientific knowledge production. 
However, recent reports hint at a potential decline in collaboration between these two giants, even amidst the backdrop of advancing global convergence. 
Understanding the intricate interplay between cooperation and disparity within the US--China relationship is vital for both academia and policy leaders, as it provides invaluable insights into the potential future trajectory of global science collaboration. 
Despite its significance, there remains a noticeable dearth of quantitative evidence that adequately encapsulates the dynamism across disciplines and over time. 
To bridge this knowledge gap, this study delves into the evolving landscape of interaction between the US and China over recent decades. 
This investigation employs two approaches, one based on paper identifiers and the other on researcher identifiers, both obtained from bibliometric data sourced from OpenAlex. 
From both approaches, our findings unveil the unique and dynamic nature of the US--China relationship, characterised by a collaboration pattern initially marked by rapid convergence, followed by a recent phase of divergence.

\vspace{0.8cm}

\vfill

\thispagestyle{empty}
\setcounter{page}{1}
\setcounter{footnote}{0}
\setcounter{figure}{0}
\setcounter{table}{0}
\setcounter{equation}{0}

\setlength{\skip\footins}{10mm}
\setlength{\footnotesep}{4mm}

\vspace{-1.6cm}

\newpage
\renewcommand{\thefootnote}{\arabic{footnote}}

\setlength{\skip\footins}{10mm}
\setlength{\footnotesep}{4mm}

\let\oldheadrule\headrule
\renewcommand{\headrule}{\color{VioletRed}\oldheadrule}

\pagestyle{fancy}
\fancyhead[LE,RO]{\textcolor{VioletRed}{\footnotesize{\textsf{\leftmark}}}}
\fancyhead[RE,LO]{}
\fancyfoot[RE,LO]{\color[rgb]{0.04, 0.73, 0.71}{}}
\fancyfoot[LE,RO]{\scriptsize{\textbf{\textsf{\thepage}}}}
\fancyfoot[C]{}
\thispagestyle{empty}

\newpage
\tableofcontents
\vfill

\begin{mdframed}[linecolor=magenta]
\begin{tabular}{l}
\begin{minipage}{0.09\hsize}
\hspace{-2mm}\includegraphics[width=1.3cm,clip]{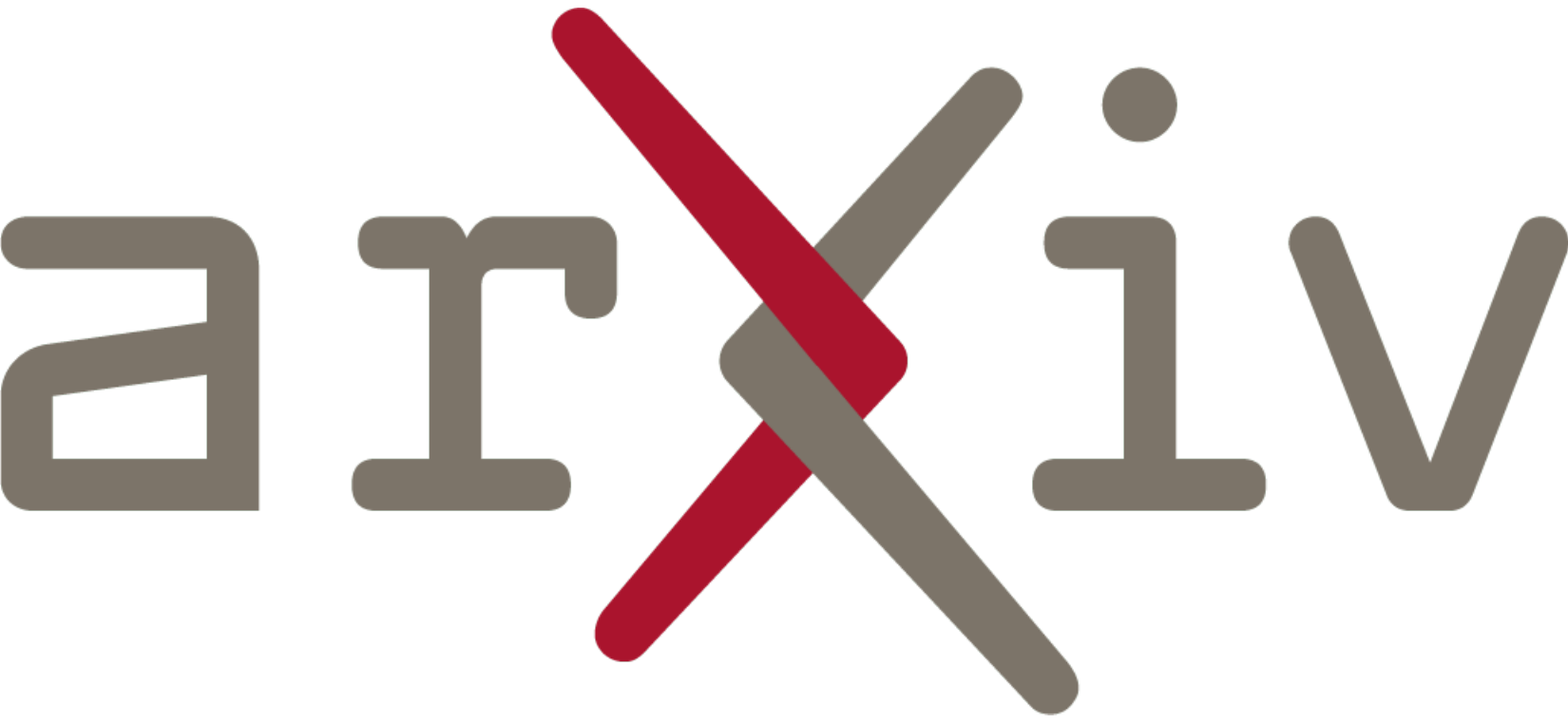}
\end{minipage}
\begin{minipage}{0.9\hsize}
\setstretch{0.84}
\textcolor{magenta}{\footnotesize\textrm{%
This version is a preprint of a paper authored by K.~Kitajima and K.~Okamura, titled \textit{\q{\mtitle}}, which has been posted on arXiv. 
The Supplementary Materials that accompany this manuscript have been seamlessly integrated into this single file.}}
\end{minipage}
\end{tabular}
\end{mdframed}

\newpage
\section{Introduction\label{sec:Introduction}}

Throughout history, the realm of science and technology (S\&T) has advanced through a delicate balance of competition and collaboration. 
At various levels, be it individuals, teams, organisations, countries or economic regions, the interplay of competition and collaboration across borders has yielded unique value, propelling modern civilisation forward. 
While S\&T has brought numerous benefits, it has also posed various threats, risks, and disruptions. 
These challenges have been overcome through cooperation at different academic levels and coordinated efforts led by governments.
Consequently, in contemporary times, the pace of scientific and technological advancement and its global impact are accelerating on a daily basis \cite{Liu23a, Fortunato18}. 
The primary drivers behind this acceleration are the rapid progress of the information society and digital platforms in recent years. 
We now stand on the threshold of an era where a single scientific discovery or technological innovation can exert a profound influence on the entire world.

Throughout this progression, not all S\&T domains have followed the same developmental trajectory. 
This holds true for international research collaboration as well, which has undergone substantial transformations across various disciplines and over time. 
Numerous factors have contributed to these transformations, including the demands and needs of the era, the unique characteristics of specific fields, the research and development (R\&D) capabilities of various institutions, national objectives and missions, international strategies, their connections to national security and industrial policies, as well as the emergence of abundant S\&T talent. 
These factors have intertwined in complex ways, collectively shaping the current scientific and technological landscape and the enterprises it propels forward.

The era where only a few prodigious talents could shape an epoch is drawing to a close. 
Today, we find ourselves firmly entrenched in an information society where vast volumes of data flow through the internet, heralding the era of data-driven and AI-empowered team science \cite{NRC15, OECD15}.
As we contemplate the multifaceted and international history that S\&T and society have traversed, we unearth valuable lessons from the history and dynamics of competition and collaboration, especially at the national level. 
These lessons bear great significance for the future.
This forms the core motivation for researchers to persist in analysing international collaborative research spanning various disciplines, armed with modern technologies and resources. 
Simultaneously, it furnishes policymakers with crucial insights and recommendations \cite{Whetsell23, Kwiek21, NSB-NSF22, OECD17, Dong17, OECD16}.
In the midst of the irreversible surge of Open Science \cite{Miedema22, Burgelman19}, the scope and depth of the field often referred to as the \q{Science of Science} \cite{Liu23a, Lin23b, Fortunato18, Zeng17} are also expanding at an exponential pace.

Efforts to analyse and visualise international cooperation in the realm of S\&T have been extensive. 
Some studies have focused on specific countries or economic regions, while others have conducted in-depth investigations into particular research disciplines---there are too many references to individual countries and disciplines to list. 
As a result, previous studies have revealed that the trend of international collaborative research, when viewed globally, has been steadily deepening over the past few decades, including recent years \cite{Okamura23b,NSB-NSF22, Kwiek21, OECD17, Dong17}. 
This trend can aptly be described as a \textit{\q{Shrinking World}} in the context of international collaborative research.
A recent paper by \citek{Okamura23} provided a global-scale quantitative demonstration of this concept. 
In that study, hierarchical clustering based on international collaboration relationships among countries was conducted, and the clustering of countries was analysed by examining the heights of branch points in the dendrogram. 
This analysis confirmed that the world has consistently been shrinking over the past half-century across various disciplines.

\begin{figure}[!t]
\centering
\vspace{-0.5cm}
\includegraphics[align=c, scale=1.25, vmargin=0mm]{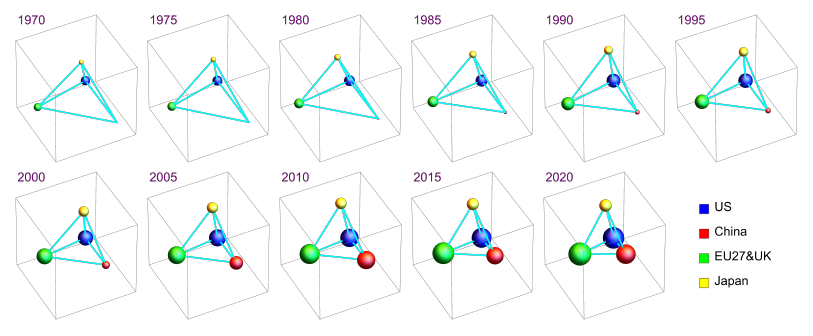}
\caption{\textbf{Changes in the mutual distance between the four parties over time.}
The four parties included in the analysis are the US, China, EU27\&UK and Japan. 
The changes over the period of 1970 to 2021 are displayed in five-year intervals. 
Each coloured sphere's volume is proportional to the number of scientific publications produced by each party, and the sizes can be compared within and across snapshots. 
The distance between the centres of each sphere represents the level of closeness in international collaboration between the corresponding parties, with shorter distances indicating closer collaboration. 
The calculation methodology for these distances is described in Section \ref{sec:Methods}.
}
\label{fig:Tetra}
\end{figure}

Reflecting on the past few decades, it becomes particularly intriguing to narrow our focus to five key players: the US, China, EU27, the UK and Japan. 
Figure \ref{fig:Tetra} quantifies and visually represents the distances between these parties over the past half-century, with a detailed methodology outlined in Section \ref{sec:Methods}. 
In this context, EU27 and the UK are regarded as a single combined entity.
The size of each sphere corresponds to the volume of scientific output associated with each of these parties, and this output has consistently expanded over the years. Simultaneously, the lengths of the edges connecting these parties, symbolising the distances between them, have markedly reduced. 
Consequently, as time has progressed, while global scientific output has dramatically increased, the volume of the overall tetrahedron has shrunk. 
This phenomenon effectively encapsulates the notion of a \textit{\q{Shrinking World}}.

In terms of knowledge and intelligence production in the realm of S\&T at the national level, the US historically held a dominant position as the sole superpower. 
This was true across numerous disciplines until roughly a decade or fifteen years ago. 
However, recent years have witnessed substantial transformations in the global landscape of scientific and technological competition and collaboration. 
The primary catalyst for this transformation has been the remarkable ascent of China over the past two decades \cite{Okamura23, Okamura23b, NatureIndex23a, NatureIndex23b, NSB-NSF22}, as evident in Figure \ref{fig:Tetra}.

When contemplating the future development of S\&T and the prospects of international collaboration, it is not challenging to envision that the prevailing trend will continue in the foreseeable future. 
In this scenario, a critical boundary condition that will influence S\&T policy decisions, similar to the conditions shaping the global economy, will be the relationship between the US and China \cite{RAND23,Toney21,Zhao19}. 
It is widely acknowledged that both countries have rapidly converged in terms of international collaborative research across various disciplines over the past few decades \cite{Okamura23,Okamura23b,NSB-NSF22}.
However, several prior studies have indicated a possible decline in collaboration between these two giants or a deterioration in their cooperative relationship, particularly since around 2019 \cite{Okamura23,Aghion23,Xie23,Jia22,VanNoorden22}.
The \q{China Initiative}, initiated by the US Department of Justice in November 2018 to address the national security threat posed by China and concluded in February 2022, is often cited as a significant factor underlying these developments \cite{Mervis23, Gilbert23}.

If the production of human knowledge and intelligence were confined to specific clusters of countries with less effective cross-border collaboration than possible, it would inevitably heighten risks for the global economy and civilised society. 
This is because, in such scenarios, addressing global challenges like preventing pandemics, bolstering resilience against natural disasters, establishing a decarbonised society, and cultivating a world that embraces diverse perspectives would be less efficient.
If there are signs of decoupling between the US and China from a geopolitical standpoint, it would present an opportunity for science diplomacy \cite{Flink21,Ruffini20,Gluckman17}, which must be grounded in modern understanding and realism. 
Promoting international collaboration based on research integrity and employing diplomatic efforts when necessary is a significant responsibility for all modern S\&T leaders.

To provide S\&T policymakers, diplomats, relevant researchers and practitioners with valuable evidence for policy deliberation, this paper analyses and visualises the changes in the relationships between major scientific powerhouses over the past decades, specifically the US, China, EU27, the UK and Japan, across various research disciplines. 
By doing so, we illustrate how the relationship between the US and China has exhibited a notably dynamic trend of convergence and divergence among these pairs of scientific powerhouses over time.

This paper is structured as follows: 
Section \ref{sec:Methods} provides an overview of the bibliometric data sourced from OpenAlex, defines the relevant concepts for the analysis, and outlines the methodology employed for analysis and visualisation. 
Section \ref{sec:Results} presents the key analysis findings.
Quantitative evidence of changes in the distance or depth of interaction between countries over the past decades is derived by examining both approaches focusing on work (paper) and author (researcher) identifiers within the OpenAlex database.
Finally, Section \ref{sec:Discussion} offers a summary, conclusions and discussions of the results.

\section{Methods\label{sec:Methods}}

This section provides a description of the data used in this study. 
All data were analysed and visualised with Python software (version 3.6), STATA/IC software (version 13; StataCorp LP, TX, USA) and Mathematica (version 13.2; Wolfram Research, Champaign, IL, USA).

\subsection{Data preparation and implementation}

\paragraph{Data source.}

The data used in this study was obtained through the \href{https://docs.openalex.org/api/}{OpenAlex API} \cite{Priem22}, an Open Bibliometrics platform launched in 2022 as a replacement for Microsoft Academic Graph (MAG) \cite{Sinha15}. 
The data was retrieved on 19th May 2023. 
OpenAlex collects information on various types of scientific publications, including journal articles, non-journal articles, preprints, conference papers, books, theses and datasets. 
These publications are collectively referred to as \q{works}, and OpenAlex indexes over 240 million of them.

The advantages and potential drawbacks of utilising OpenAlex data are extensively discussed in \citek{Okamura23}, which is pertinent to the present study.
Regarding the advantages, this paper emphasises the timely and swift accessibility of a substantial volume of freely available data. 
This accessibility streamlines the identification of broad bibliometric trends, proving particularly valuable for policy-related objectives.
One potential drawback is that the data's quality may not be on par with commercial databases. 
Nevertheless, in the context of policy objectives, the advantages mentioned earlier often outweigh this limitation. 
Policymakers frequently require current data that can be accessed, analysed, and visualised promptly as needed.

\paragraph{R\&D disciplines.}

To ensure meaningful policy implications for international research collaboration, it is important to employ an appropriate classification scheme for various R\&D disciplines. 
OpenAlex has devised a classification scheme referred to as the \q{concepts}, which encompasses field categories at various levels of granularity, spanning from level 0 to level 6.\endnote{The complete list of the \q{concepts} can be found at \url{https://api.openalex.org/concepts} (accessed 29th August 12 2023).} 
Higher levels within this scheme denote more specific concepts, while lower levels encompass broader ones. 
These concepts serve as a framework for categorising research areas.

In this study, we specifically focus on the 19 level-0 concepts from the OpenAlex classification. 
These concepts cover a wide range of disciplines, including humanities and social sciences (HSS) as well as natural sciences. 
The level-0 concepts we consider are: 
{\math}, {\comp}, {\geog}, {\bio}, {\phys}, {\chem}, {\engi}, {\mate}, {\envi}, {\med}, {\psy}, {\biz}, {\econ}, {\soc}, {\pol}, {\geol}, {\hist}, {\phil} and {\art}.
By including these diverse disciplines of study, we aim to capture a comprehensive view of international research collaboration across different domains.

\paragraph{\q{Nationality} of works and the counting method.}

To streamline the counting process, we use the term \q{work of nationality X} to refer to the work produced by contributors from institutions in country X.
In cases where contributors from both country X and country Y collaborate on a work, the work is considered to have dual nationality, thus being counted as both a work of nationality X and a work of nationality Y.
While we use the terms \q{country} or \q{nationality} in this context, it is important to note that the discussion in this paper is equally applicable to situations where it encompasses a group of countries, such as EU27.

This counting method allows works to have multiple nationalities in our analysis.
To count the number of works for each country, we employ a binary counting method based on nationality, as described in \citek{Okamura23}. 
In this method, if a work has dual nationality (X and Y), it is counted as one work output for each country. 
Even if there are multiple contributors from country X, the work is only counted as one in the production volume for country X. 
If the country information is unknown for all contributors of a particular work, it is categorised as \q{unknown} and excluded from the analysis conducted in this study.

\paragraph{Countries.}

This paper focuses on five key parties: the US, China, EU27, the UK and Japan. 
These countries were selected based on their significant contributions to works production across all disciplines of science from 1971 to 2020.
When excluding works with unknown nationality, the top five countries in terms of work production are as follows: the US (24,425,579), China (12,106,744), the UK (6,403,998), Germany (5,288,254) and Japan (4,768,966). 
Following closely are France (4,318,698), Canada (3,154,311), India (3,001,958) and Italy (2,766,564) (as of 12th August 2022). 
These selections align with the countries recognised as the \q{Big 5 science nations} (the US, China, Germany, the UK and Japan) in the Nature Index \cite{NatureIndex22}.
While other countries like Canada and India also present interesting insights, limiting the analysis to these five parties ensures clarity in presenting the results. 
It is worth noting that the methodology described in this paper can be applied to include additional parties or focus on different countries, maintaining its applicability and relevance.

\subsection{The `{Collaboration Distance}' approach}

There can be various approaches to quantifying bilateral distance based on coauthorship relationships between two countries.
One of the simplest approaches might suggest that the greater the number of coauthored papers, the closer the distance between the countries.
However, adopting such a simplistic approach falls short of satisfying a crucial mathematical property of distance known as the triangle inequality.
This limitation can hinder certain formal analyses, such as clustering analysis.
In the quantitative discussion of whether international research collaboration is globally \q{contracting} or if the world is \q{shrinking} in this context, it is also crucial to measure the relevant \q{distance} and the corresponding \q{volume}.
Therefore, a certain level of refinement is necessary.
It is essential that any modifications made during this refinement process are not overly complex or costly.
Given the significant interest of S\&T policymakers in cooperative and distance relationships within academia between countries, it is necessary to adopt a practical and easily interpretable method that suits their needs.
Still, the definition of distance should ideally satisfy the triangle inequality.
This is particularly significant because policymakers are always highly concerned about the existence of collaborative relationship clusters at the international level and the positioning of their own country within those clusters.
To meet all these requirements, this paper adopts the same simple set-theoretic distance method employed in \citek{Okamura23}.

Let $S_{\X|\alpha,\tau}$ denote the set of works of nationality X published in discipline $\alpha$ during a given period $\tau$. 
Then, the distance $D_{\alpha,\tau}(\X,\Y)$ between countries X and Y in the given discipline and period is defined as:
\begin{equation}\label{eq:distance}
D_{\alpha,\tau}(\X,\Y)
=1-\f{\abs{S_{\X|\alpha,\tau}\cap S_{\Y|\alpha,\tau}}}{\abs{S_{\X|\alpha,\tau}\cup S_{\Y|\alpha,\tau}}}\,.
\end{equation}
This distance metric is known as the Jaccard distance and ranges from 0 to 1. 
A value of 0 indicates that X and Y are identical, while a value of 1 indicates that they are entirely distinct. 
The Jaccard distance satisfies the mathematical definition of distance in the set-theoretic sense, including the triangle inequality.
Based on the resulting $n \times n$ distance matrix $\bmt{D}_{\alpha,\tau}$ for each discipline and period, where $n$ represents the number of countries (parties), it is possible to embed the position coordinates $\bmt{x}_{i|\alpha,\tau}$ of each country $\X_{i}$ ($i=1,\,\dots,\,n$) in an $(n-1)$-dimensional Euclidean space.\endnote{The embedding process follows the procedure outlined in Appendix A.3 of \citek{Okamura23}.}
Hereafter, the distance measure (\ref{eq:distance}) is transformed through rescaling: $D\mapsto {D}_{\mathrm{resc}}\coloneqq -\ln D$, aiming to improve visual effectiveness.

\subsection{The `{Knowledge Flow}' approach}

Coauthorship relationships in works discussed above can be viewed as an \q{undirected-graph} approach, where the symmetrical relationship between two countries (parties), X and Y, is expressed through a distance function. 
However, it is equally valuable and highly relevant for S\&T policymakers, diplomats and practitioners to capture unilateral movements of researchers from one country to another, such as inflows or outflows based on affiliations.
To address this point, we also take on a \q{directed-graph} or \q{asymmetric} approach, focusing on author (researcher) identifiers instead of work (paper) identifiers. 
This direction leads us to introduce the concept of \q{knowledge flow}, which extends the conventional idea of researcher mobility. 
Below, for convenience, we consider the temporal flow on a yearly basis, but it is essential to emphasise that the concept of knowledge flow is applicable to any appropriate time unit.

Let us consider a researcher, P, affiliated with universities in both country X and country Y in a certain year $\tau$. 
This fact can be inferred from the \q{affiliation} field of P's produced work, where both universities from country X and country Y are listed. 
In the subsequent year ($\tau+1$), we assume researcher P remains affiliated with universities in both countries X and Y. 
Consequently, we interpret that academic knowledge flows between the two countries through the mediator, researcher P, across consecutive years ($\tau\to \tau+1$), implying knowledge exchange from X to Y and vice versa.
This concept of knowledge flow contrasts with the common understanding of researcher mobility, which often assumes physical relocation \cite{RAND23,Vaccario21,Sugimoto17,RAND17}. 
Instead, we focus on how researchers' affiliations change within what we term the affiliation space. 
This leads to our \q{directed-graph} or \q{asymmetric} approach, considering researcher identifiers and diverging from the \q{undirected-graph} or \q{symmetric} approach, which centres on coauthorship relationships based on work identifiers.
Now, let us consider the following year ($\tau+2$), where researcher P is exclusively affiliated with a university in country X. 
We interpret that academic knowledge flows from country Y to country X through researcher P across years ($\tau+1\to \tau+2$), but not in the opposite direction. 
The rationale behind this interpretation is rooted in the fact that researcher P is no longer affiliated with country Y, resulting in the discontinuation of any academic knowledge previously held by P while affiliated with country X from flowing to country Y through P.
Likewise, in the subsequent year ($\tau+3$), assuming that researcher P is exclusively affiliated with a university in country Z, we can infer the flow of academic knowledge from country X to country Z through researcher P over the years ($\tau+2\to \tau+3$), but not in the direction from Z to X or Y.

To facilitate a quantitative approach, let us introduce some notations. 
For each year ($\tau$), we quantify the number of knowledge flows from country $\X_{i}$ to country $\X_{j}$ in the subsequent year ($\tau+1$).
We denote this count as the $(i,j)$-th element of the $K$-matrix, referred to as $K_{\X_{i}\to\X_{j}}(\tau)$, or more concisely as $K_{ij}(\tau)$, defined for year $\tau$. 
Here, the labels $i$, $j$ and so on represent distinct countries (parties).
In accordance with this convention, the diagonal elements, $K_{ii}(\tau)$, represent the number of researchers who maintain affiliations with institutions in the country $\X_{i}$ over the years.
Since each element of the calculated $K$-matrix is influenced by researchers' overall mobility trend and the total number of works they produce, it is crucial to scale each element appropriately relative to such absolute quantities.
Our specific interest lies in the knowledge flows between different countries, which correspond to the off-diagonal elements of the $K$-matrix. 
Therefore, for each year $\tau$, we calculate the sum of the off-diagonal elements of $K$, denoted as $\Sigma_{\mathrm{off}}(K) \coloneqq \sum_{i\neq j}K_{ij}$, and utilise this off-diagonal sum to derive the off-diagonal elements of the scaled $K$-matrix, $\cK_{ij} \coloneqq K_{ij}/\Sigma_{\mathrm{off}}(K)$ ($i\neq j$).
This quantity represents our focal point, the Knowledge Flow Rate (KFR).\endnote{When illustrating the analysis results graphically in the subsequent section, the KFR index will be multiplied by 100 to express it as a percentage.}
Note that this quantity is normalised over all combinations of distinct pairs of $i$ and $j$, i.e.~$\sum_{i\neq j}\cK_{ij}=1$.
In other words, each $\cK_{ij}$ quantitatively represents the proportion of knowledge flow from $\X_{i}$ to $\X_{j}$ among all unique pairs of countries within the total knowledge flow occurring during a particular time transition.

By observing the temporal changes in $\cK_{ij}(\tau)$, we can gauge the relative intimacy of countries $\X_{i}$ and $\X_{j}$.
It is crucial to note that the $\cK$-matrix in this context is an asymmetric matrix, i.e., $\cK_{ij}\neq\cK_{ji}$ in general.
To demonstrate the implications of this characteristic, let us consider a scenario during a particular time transition from $\tau\to\tau+1$, where $\cK_{ij}>\cK_{ji}$ holds for a given $i$ and all $j$ for simplicity.
In this case, country $\X_{i}$ is suggested to experience more knowledge flowing out to other countries than knowledge flowing in from other countries during this period.
Furthermore, when focusing on a specific pair of $i$ and $j$, a situation where the value of $\cK_{ij}$ increases (or decreases) over the years indicates a relative upward (or downward) trend in the proportion of knowledge flowing from country $\X_{i}$ to country $\X_{j}$.
However, it should be noted that $\cK_{ji}$ may not necessarily follow the same trend of temporal changes.
In cases where there is a significant asymmetric knowledge absorption potential between countries $\X_{i}$ and $\X_{j}$, it is possible that $\cK_{ji}$ shows a decreasing (increasing) trend while $\cK_{ij}$ simultaneously exhibits an increasing (decreasing) trend.

The aforementioned approach might appear somewhat heuristic, but it remains effective in quantifying and analysing the flow of knowledge over time.
It proves to be a valuable tool for understanding the circulation of knowledge among researchers, as will be demonstrated subsequently.

\section{Results\label{sec:Results}}

This section provides a detailed analysis results of the two approaches introduced in the previous section: the \q{Collaboration Distance} approach and the \q{Knowledge Flow Rate (KFR)} approach. 
These methods centre around distinct identifiers, specifically work (paper) identifiers and author (researcher) identifiers. 
Interestingly, both approaches yield consistent quantitative evidence and insights from different perspectives, particularly concerning the relationship between the US and China.

\subsection{Change in the `{Collaboration Distance}'}

\paragraph{Evolution of collaboration distance by field.}

Based on the methodology described earlier, we computed and visually represented the distances (${D}$) between pairs of the five parties: the US, China, EU27, the UK and Japan, across different time periods and the 19 focused disciplines.
See Suppl.~Fig.~\ref{sfig:19F_5P}, where the vertical axis of the graph represents the bilateral distance.
Among the three parties---the US, EU27 and the UK---there have been generally similar trends in the changes of their distances across various disciplines, with some variations in absolute levels.
The distances between these parties have consistently decreased in a nearly linear manner, and they have maintained the shortest distances throughout the analysed periods.

On the contrary, the distances between China and any of the other parties have typically been greater when compared to the distances among Western countries within each time period.
Nonetheless, there has been a similar trend of distance reduction over time.
Of particular interest is the rapid decline in distance between the US and China since the beginning of the 21st century, with a steep decline in the early 2010s.
This phenomenon is noteworthy not only in natural science disciplines but also in HSS disciplines, such as {\hist}, {\phil} and {\art}.

However, since 2019, there has been a reversal in this trend, with distances tending to widen or remain relatively stable.
This trend is particularly pronounced in the natural sciences, exhibiting an observable \q{time-reversed J-curve} pattern.
This observation aligns with a phenomenon reported in a prior study \cite{Okamura23} using a different set of field categories (specifically, some level-1-related natural science concepts in OpenAlex).
Although the curvature of distance convergence is less prominent in the HSS disciplines compared to the natural science disciplines, there is still a tendency for deceleration.
In contrast to the natural sciences, particular disciplines within humanity are known to often require more time for international collaborative research to reach maturity \cite{Kwiek21}. 
Therefore, there is a possibility that, with some time delay, we may also witness an increasing divergence between the US and China in these disciplines.

Although there are other intriguing implications in the relationships between various pairs of parties, we will not explore them here. 
This paper specifically concentrates on the US--China relationship and underscores its relative distinctions in comparison to other pairs.

\begin{figure}[!t]
\centering
\vspace{-0.5cm}
\includegraphics[align=c, scale=0.8, vmargin=0mm]{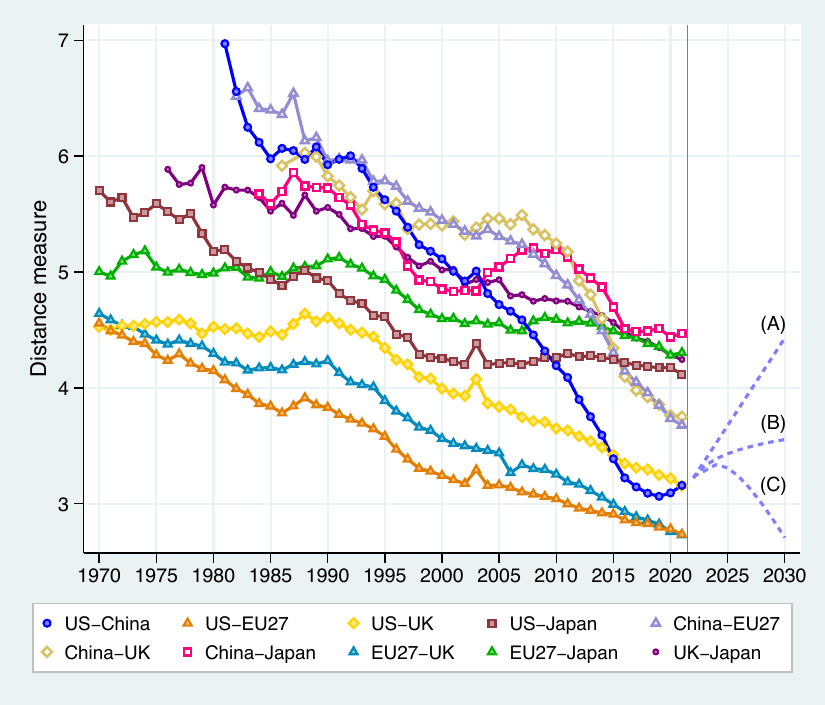}
\caption{\textbf{Change in the Collaboration Distance among the scientific powerhouses in natural sciences over time.}
In addition to the observed values for 1990--2021, simulation results until 2030 based on specific scenarios (A--C) are shown for the US--China pair.}
\label{fig:line_dist_natall_5P}
\end{figure}

\paragraph{\textit{`Shrinking (and-possibly-Polarising) World'}.}

Figure \ref{fig:line_dist_natall_5P} presents the temporal evolution of pairwise distances in the natural science disciplines for the same five parties mentioned earlier (the US, China, EU27, the UK and Japan).
The natural science disciplines encompass all the level-0 categories of natural science concepts in OpenAlex, including {\math}, {\comp}, {\geog}, {\bio}, {\phys}, {\chem}, {\engi}, {\mate}, {\envi} and {\med}.
The figure provides an analysis and visualisation of the changing distances between each pair of parties from 1970 onwards.
The overall trend observed in the figure represents an average depiction of the trends observed in each field, including the \q{time-reversed J-curve} characteristic observed in the US--China relationship.

Regarding the distances between the US and China, the data from 1970 to 2021 is based on actual observed values, while the graph from 2022 to 2030 is a simulation result based on certain assumptions.
Three scenarios (A, B and C) are presented.
Scenario A depicts a situation where the distance between the two countries diverges at the same rate as it had been rapidly converging, almost like rewinding a clock.
In other words, it represents a scenario where the distance between the two countries widens at the same pace as it had been narrowing.
By contrast, Scenario B shows a moderate progression where the distance continues to increase but at a slightly slower growth rate.
Lastly, Scenario C suggests that although the distance between the US and China will continue to increase for the next few years, it will soon transition back to a decreasing trend.

\begin{figure}[!tp]
\centering
\vspace{-0.5cm}
\hspace{-0.5eM}\includegraphics[align=c, scale=0.5, vmargin=0mm]{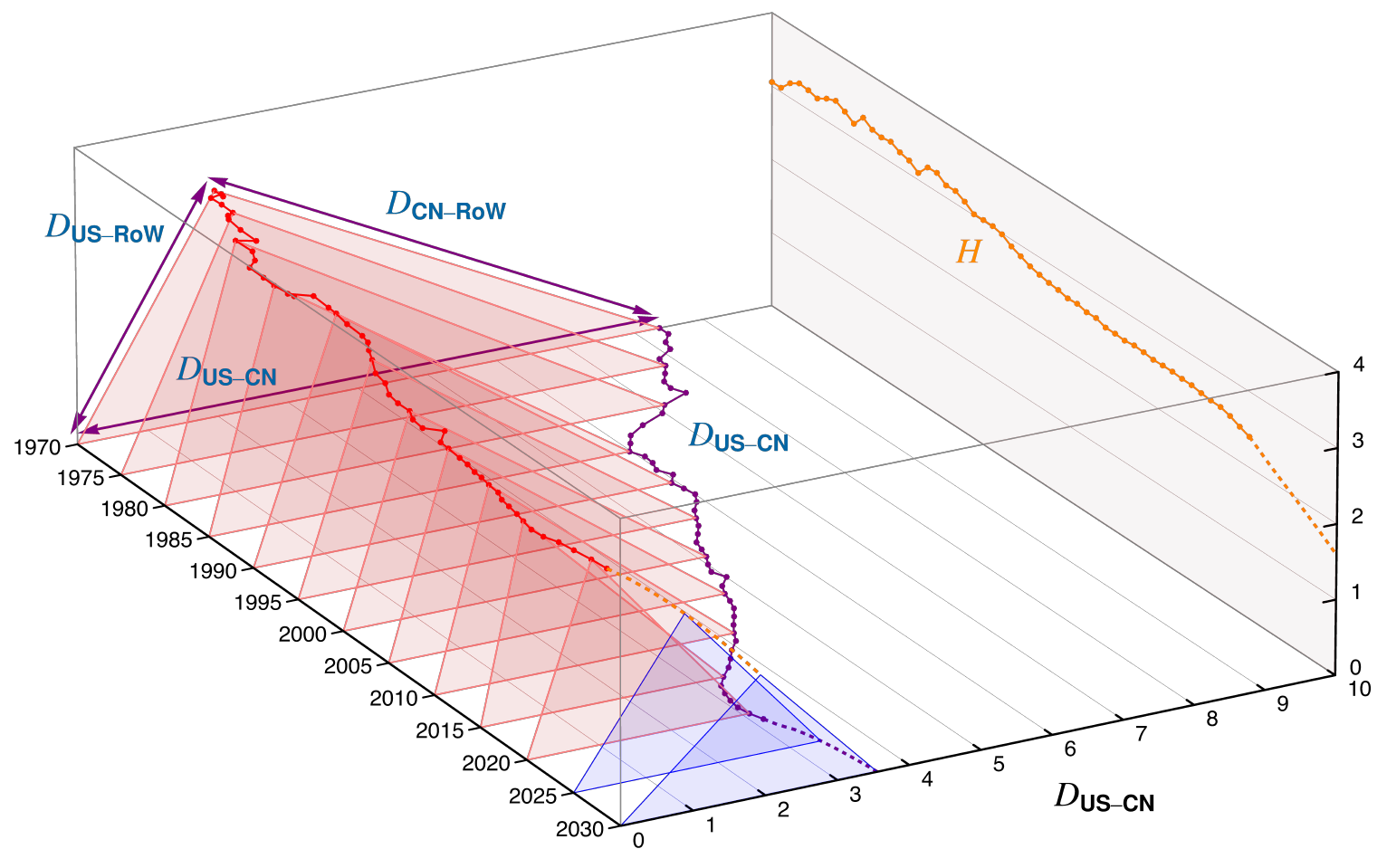}
\caption{\textbf{`Shrinking (and-possibly-Polarising) World'.}
The data prior to 2021 represents actual values obtained from observational data, while the data from 2022 onwards illustrates results derived from simulations based on Scenario B.
The abbreviation \q{RoW} stands for \q{Rest of the World's top 50}, comprising the top 50 countries, following the US and China in terms of work production.
The decreasing trend in the area of the triangle formed by the US, China and RoW signifies the visualisation of the \textit{\q{Shrinking World}}.
}
\label{fig:SPWorld}
\end{figure}

Figure \ref{fig:SPWorld} illustrates the \textit{\q{Shrinking World}} of science collaboration in the natural science disciplines over the past 50 years, focusing specifically on the US and China and simplifying the world into three parties: the US, China and the Rest of the World (RoW).
The RoW represents the collective scientific output of the top 50 countries, following the US and China in terms of work production.\endnote{The RoW includes the UK, Japan, Germany, France, Canada, India, Italy, Australia, Spain, Brazil, Russia, South Korea, Netherlands, Poland, Switzerland, Indonesia, Sweden, Iran, Taiwan, Belgium, Turkey, Denmark, Israel, Mexico, Austria, Norway, Finland, South Africa, Czechia, Portugal, Greece, Malaysia, Singapore, Egypt, New Zealand, Argentina, Saudi Arabia, Ukraine, Ireland, Hungary, Pakistan, Thailand, Colombia, Chile, Romania, Nigeria, Slovakia, Croatia, Serbia and the Philippines.}
The visual representation depicts the world as a triangle, with snapshots provided at five-year intervals.
The base of the triangle is formed by two points representing the US and China, and the length of the line segment connecting these two points represents the distance between the US and China ($D_{\mathrm{US-CN}}$).
The apex of the triangle represents the RoW, and the lengths of the line segments connecting the apex to the endpoints of the base ($D_{\mathrm{US-RoW}}$, $D_{\mathrm{CN-RoW}}$) represent the distances between the RoW and the US and China, respectively.
The height ($H$) of the triangle is also plotted along the vertical axis.

From the figure, it is evident that over the past half-century, the area of the \q{US--China--RoW Triangle} has rapidly decreased, with some fluctuations, as indicated by the decreasing lengths of all three sides (as shown in Fig.~\ref{fig:line_dist_natall_5P}).
The widening distance between the US and China from 2019 to 2022 is reflected in the expansion of the base ($D_{\mathrm{US-CN}}$) of the triangle.
The plot from 2022 to 2030 in Fig.~\ref{fig:SPWorld} is a simulation based on Scenario B from Fig.~\ref{fig:line_dist_natall_5P}, provided for reference.
While $D_{\mathrm{US-RoW}}$, $D_{\mathrm{CN-RoW}}$, and the height $H$ decrease almost linearly, $D_{\mathrm{US-CN}}$ shows an increasing trend, resulting in the flattening and stretching of the triangle.
If Scenario A were adopted, the triangle would become even more elongated.
Conversely, if Scenario C were adopted, all three sides would become shorter, resulting in a smaller triangle with a smaller perimeter and area.
The actual realisation of Scenarios A to C in the future US--China relationship remains uncertain.
Nevertheless, this triangular diagram remains an intriguing focal point, potentially offering a bibliometric perspective on the US--China relationship. 
The discussions in this section will be beneficial for future deliberations on international collaboration in the policy arena.

\subsection{Change in the `{Knowledge Flow}' trend}

We also conducted an analysis of the Knowledge Flow Rate (KFR) for each discipline at level-0 in OpenAlex. 
Our analysis focused on a group of the top 199 researchers in terms of work production each year.\endnote{When using OpenAlex's API with the \q{\texttt{group\_by=authorships.author.id}} filter, a maximum of 200 data points can be obtained (as of August 2023), one of which belongs to authors with unknown identifiers. 
Consequently, the maximum number of obtainable author data through this method was 199.}
In contrast to the broader sample used in the Collaboration Distance analysis, our study on the KFR necessitated a narrower focus on the period between 2000 and 2021. 
This particular time frame was chosen due to the stabilisation of the sum of off-diagonal elements ($\Sigma_{\mathrm{off}}$), ensuring the reliability of our analysis. 
Cases where $\Sigma_{\mathrm{off}}$ was less than 10 for a given year were excluded from the calculation of the KFR index.
Similar to the Collaboration Distance analysis detailed in the preceding section, our investigation also centred on the identical set of five parties: the US, China, EU27, the UK and Japan.

The analysis results for each discipline are summarised in Suppl.~Fig.~\ref{sfig:kflow_all}. 
A key observation emerges: over the past two decades, two KFR levels between the US and China (i.e.~$\cK_{\mathrm{US}\to\mathrm{CN}}$ and $\cK_{\mathrm{CN}\to\mathrm{US}}$) consistently outperformed other pairs, a trend evident across all disciplines. 
Additionally, an intriguing trend becomes apparent across various natural science disciplines. 
Since the transition from 2018 to 2019, there has been a notable decrease in the KFR from China to the US. 
This shift implies that China is rerouting the outflow of knowledge away from the US towards other entities. 
This observation aligns effectively with the results obtained from the Collaboration Distance analysis (Suppl.~Fig.~\ref{sfig:19F_5P}).
Furthermore, the relatively subdued trend observed in the HSS disciplines, as opposed to the natural sciences, in terms of the decreasing trend post-2018, aligns with the findings from our Collaboration Distance analysis as well.

\begin{figure}[!t]
\centering
\vspace{-0.5cm}
\includegraphics[align=c, scale=0.7, vmargin=0mm]{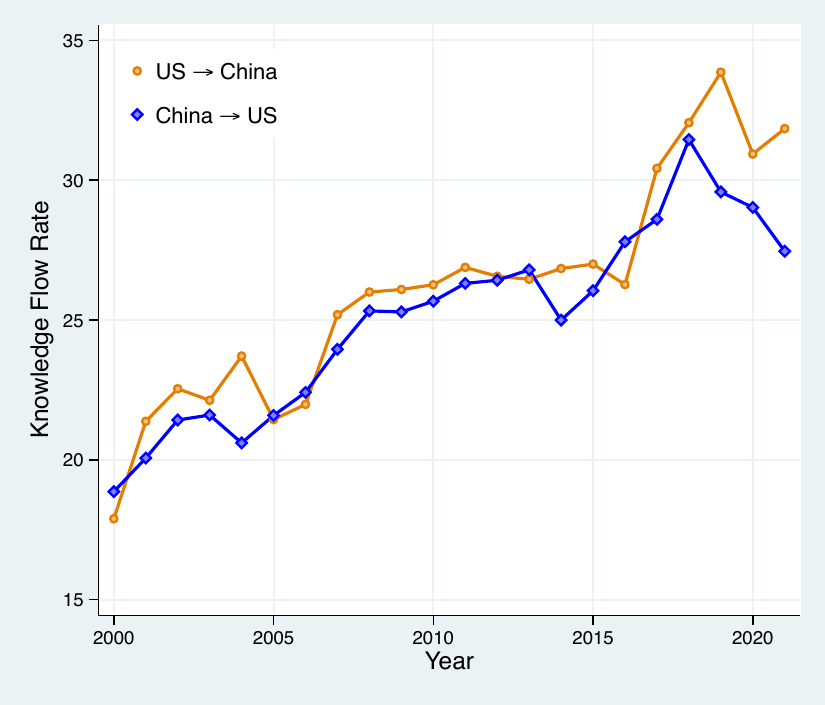}
\caption{\textbf{Change in the Knowledge Flow Rate between the US and China in natural sciences over time.}}
\label{fig:kflow_natall_USCN}
\end{figure}

Given the consistent trends across various natural science disciplines, it is valuable to amalgamate these disciplines and derive an average representation of knowledge flow within the domain of natural sciences. 
Such an analysis can provide insights into overall trends across all level-0 natural science disciplines. 
To avoid overwhelming with data, Fig.~\ref{fig:kflow_natall_USCN} focuses on the average KFR between the US and China, while Suppl.~Fig.~\ref{sfig:kflow_natall_Others} presents the remaining results.
In Fig.~\ref{fig:kflow_natall_USCN}, the KFRs within natural science disciplines between the US and China have shown a marked upward trend since 2000. 
Specifically, both $\cK_{\mathrm{US}\to\mathrm{CN}}$ and $\cK_{\mathrm{CN}\to\mathrm{US}}$ indicators, which were below 20\% of total knowledge flow among the five parties in 2001, exceeded 30\% by 2018. 
This highlights a steady growth in knowledge exchange between the US and China facilitated by productive researchers, surpassing exchanges between other pairs of countries. 
This trend reaffirms the deepening connection over the past two decades. 
However, as mentioned earlier, this trend ceased to persist after the transition from 2018 to 2019.

Noteworthily, the asymmetrical nature of the KFR matrix, where $\cK_{ij}\neq \cK_{ji}$, offers the potential for more detailed insights compared to Collaboration Distance analysis. 
As an example, during the period from 2018 to 2021, many natural science disciplines demonstrated a decline in $\cK_{\mathrm{CN}\to\mathrm{US}}$, whereas $\cK_{\mathrm{US}\to\mathrm{CN}}$ displayed a distinct zigzag pattern: it increased from 2018 to 2019, decreased from 2019 to 2020, and then increased again from 2020 to 2021.
This phenomenon is consistently observed across various natural science disciplines (Suppl.~Fig.~\ref{sfig:kflow_all}), especially in {\math}, {\comp}, {\bio}, {\phys}, {\chem}, {\envi} and {\med}. 
The asymmetrical trend of knowledge flow between the US and China, with its significance on a global scale, also holds intriguing implications across various domains.

\section{Summary, conclusions and discussions\label{sec:Discussion}}

\paragraph{Summary and conclusions.}

This paper aimed to furnish S\&T policymakers, diplomats, relevant researchers and practitioners with substantial quantitative evidence regarding the evolution and current status of international research collaboration.
This examination focused on the US, China, EU27, the UK and Japan, encompassing various research domains.
The following quantitative evidence was presented:

\begin{enumerate}
\setlength\itemsep{-0.2em}
\item The world has consistently experienced a shrinking trend attributed to the deepening of international collaborative research---the \textit{\q{Shrinking World}}.
\item Over the past few decades, there has been a rapid decrease in the distance between the US and China when compared to other pairs of countries.
\item However, starting in 2019, signs of divergence have emerged between the US and China.
\item This trend of divergence is not limited to any specific field; instead, it spans across a wide range of disciplines. 
Notably, it is particularly prominent in the natural sciences and relatively less pronounced in the Humanities and Social Sciences (HSS).
\item The observed divergence between the US and China is unique to their relationship and is not observed in other pairs of parties.
\end{enumerate}

\noindent
The above findings were derived from an analysis of Collaboration Distance, computed based on coauthorship relationships between each pair of countries (parties).
To offer supplementary validation from a different angle, this paper also centred on a group of highly productive authors. 
The investigation analysed the Knowledge Flow Rate (KFR) established among the five parties while investigating how the information about authors' affiliations changed over time.
The observed outcomes aligned with the earlier findings, thereby reinforcing their reliability.

Considering the fourth and fifth characteristics outlined above, it becomes evident that the observed divergence between the US and China cannot be solely attributed to universal factors like the impact of COVID-19, affecting all countries.
Had these factors been the sole drivers, we would anticipate similar effects in other pairs of countries.
Furthermore, this divergence does not solely stem from the characteristics of specific research disciplines; rather, it represents unique and exceptional dynamics between the US and China.
Hence, it is reasonable to associate this divergence with underlying geopolitical factors that significantly influence the dynamics of international research collaboration.
These findings provide a valuable complement to the preliminary discussion presented in \citek{Okamura23}, where the concept of a \textit{\q{Polarising World}} of science collaboration was suggested.

\paragraph{Implications.}

While there is an increasing body of literature addressing the recent divergence between the US and China and the decline in research collaboration between them, the number of quantitative studies illustrating these trends remains limited.
One notable study by \citek{VanNoorden22}, which utilised data from the Scopus database, demonstrated a substantial decrease of over 20\% in the number of scholars affiliating with both China and the US in their research papers after 2019.
The study also highlighted a decrease in research papers resulting from collaboration between the US and Chinese authors.
Another study by \citek{Xie23}, employing data from the MAG database, observed that Chinese scientists based in the US have been increasingly inclined to leave the country, with a steady rise in return migration.
Additionally, these researchers exhibited a lower motivation to apply for US federal government grants. 
Furthermore, a study by \citek{Jia22} focused on the life sciences field, utilising data from PubMed and Dimensions, and underscored the adverse effects on US-based scientists collaborating with Chinese coauthors. 
This impact was particularly evident in terms of diminished perceived influence, measured by the number of citations their published papers received.
Complementing these findings, an additional study by \citek{Aghion23}, based on the Scopus database, demonstrated that the China Initiative had a noteworthy negative impact on the average quality of publications (measured by citation counts) and on the coauthors of Chinese researchers with prior collaborations in the US (measured by the average H-index).

The current paper utilised data from OpenAlex, an Open Bibliometrics platform, to offer consistent insights that align with the aforementioned studies highlighting the declining trend in the US--China relationship in recent years.
Our analysis holds significance in its ability to capture the convergence--divergence dynamics across a diverse range of research disciplines.
Notably, we were able to provide quantitative evidence using two distinct methodologies: the Collaboration Distance approach based on work (paper) identifiers and the KFR approach based on author (researcher) identifiers.

\paragraph{Limitations and future perspectives.}

Finally, we discuss the limitations of this study and explore potential future perspectives. 
Concerning our adoption of the OpenAlex dataset, discussions surrounding the limitations and the validity of the conclusions drawn in this paper mirror those presented in \citek{Okamura23}. 
As outlined in Section \ref{sec:Methods}, OpenAlex offers the advantage of providing extensive data in a preferred format with ongoing enhancements in service quality. 
However, commercial databases might excel in terms of data quality in specific aspects. 
A recent study by \citek{Nguyen23} has examined the data quality for the analysis of international research collaboration across various bibliometric databases, including MAG, Web of Science, Dimensions and the ACM Digital Library. 
Considering such perspectives, it would be insightful for forthcoming research to compare the outcomes of this study with findings derived from other databases.

In addition, the validity of both the Collaboration Distance and the KFR index employed in this paper, as representations of bilateral relations and temporal shifts within academia, would raise a fundamental inquiry. 
The notion of a \q{close relationship} among academic entities can be interpreted in various ways, prompting the exploration of supplementary methods to scrutinise bilateral connections. 
In conjunction with the undirected graph approach of coauthorship analysis, there lies a particular interest in conducting a more comprehensive inquiry using a directed graph-like technique that accentuates researcher-level analysis, as preliminarily demonstrated in this study. 
To refine the precision of knowledge flow analysis, improving the original data source's accuracy is undeniably crucial. 
Moreover, broadening the scope to encompass a greater number of researchers and countries (parties), thereby increasing the sample size, would contribute to more robust outcomes.
In this study, the analysis utilised the OpenAlex API, which imposed a maximum cap of 199 authors for simultaneous examination. 
Nevertheless, in forthcoming endeavours, implementing a similar methodology on a larger and more comprehensive database would be invaluable in corroborating the findings established in this paper.

Furthermore, such an approach could illuminate the underlying factors contributing to the observed results, particularly the intriguing phenomenon of the \q{time-reversed J-curve} between the US and China. 
Scrutinising which of the three scenarios (A--C) outlined in Section \ref{sec:Results} closely aligns and delving into the reasons behind it are pivotal for the future contemplation of global-scale international research collaboration. 
Such meticulous analysis would provide invaluable insights for both policymakers and the academic community.

\vspace{2mm}
\paragraph{\textbf{Acknowledgements.}}
The authors would like to thank Hitoshi Koshiba for his valuable comments.
The views and conclusions contained herein are those of the authors and should not be interpreted as necessarily representing the official policies or endorsements, either expressed or implied, of any of the organisations with which the authors are currently or have been affiliated in the past.

\vspace{0em}
\paragraph{\textbf{Author Contributions.}}
\textrm{KK}: 
Software, Validation, Investigation, Writing (Review \& Editing) and Visualisation.
\textrm{KO}: 
Conceptualisation, Methodology, Software, Validation, Formal analysis, Investigation, Data curation, Writing (Original Draft, Review \& Editing), Visualisation, Supervision and Project administration.

\vspace{0em}
\paragraph{\textbf{Competing Interests.}}
The authors have no competing interests.

\vspace{0em}
\paragraph{\textbf{Funding Information.}}
The authors did not receive any funding for this research.

\vspace{0em}
\paragraph{\textbf{Data Availability.}}
The datasets and figures generated and/or analysed during this study will be made available in accordance with the policies of the journal or other publication outlets where it will eventually appear.

\vspace{0cm}
\theendnotes
\addcontentsline{toc}{section}{Notes}


\bibliographystyle{apalike-imp}
\addcontentsline{toc}{section}{References}
\setlength{\bibsep}{0\baselineskip plus 0.2\baselineskip}
\renewcommand*{\bibfont}{\footnotesize}


\clearpage
\addcontentsline{toc}{section}{Supplementary Materials}

\renewcommand{\thesubsection}{Appendix\,~\Alph{subsection}}
\renewcommand{\thesubsubsection}{\Alph{subsection}.\,\arabic{subsubsection}}

\pagestyle{fancy}
\fancyhead[LE,RO]{\textcolor{orange}{\footnotesize{\textsf{SUPPLEMENTARY MATERIALS}}}}
\fancyhead[RE,LO]{}
\fancyfoot[RE,LO]{\color[rgb]{0.04, 0.73, 0.71}{}}
\fancyfoot[LE,RO]{\scriptsize{\textbf{\textsf{\thepage}}}}
\fancyfoot[C]{}

\renewcommand{\thefigure}{S\arabic{figure}}
\renewcommand{\thetable}{S\arabic{table}}
\renewcommand{\theequation}{S\arabic{equation}}
\setcounter{section}{0}
\setcounter{subsection}{0}
\setcounter{figure}{0}
\setcounter{table}{0}
\setcounter{equation}{0}

\renewcommand{\headrule}{\color{orange}\oldheadrule}

\addcontentsline{toc}{subsection}{Supplementary Figures}

\begin{figure}[H]
\centering
\vspace{-0.5cm}
\includegraphics[align=c, scale=1.18, vmargin=0mm]{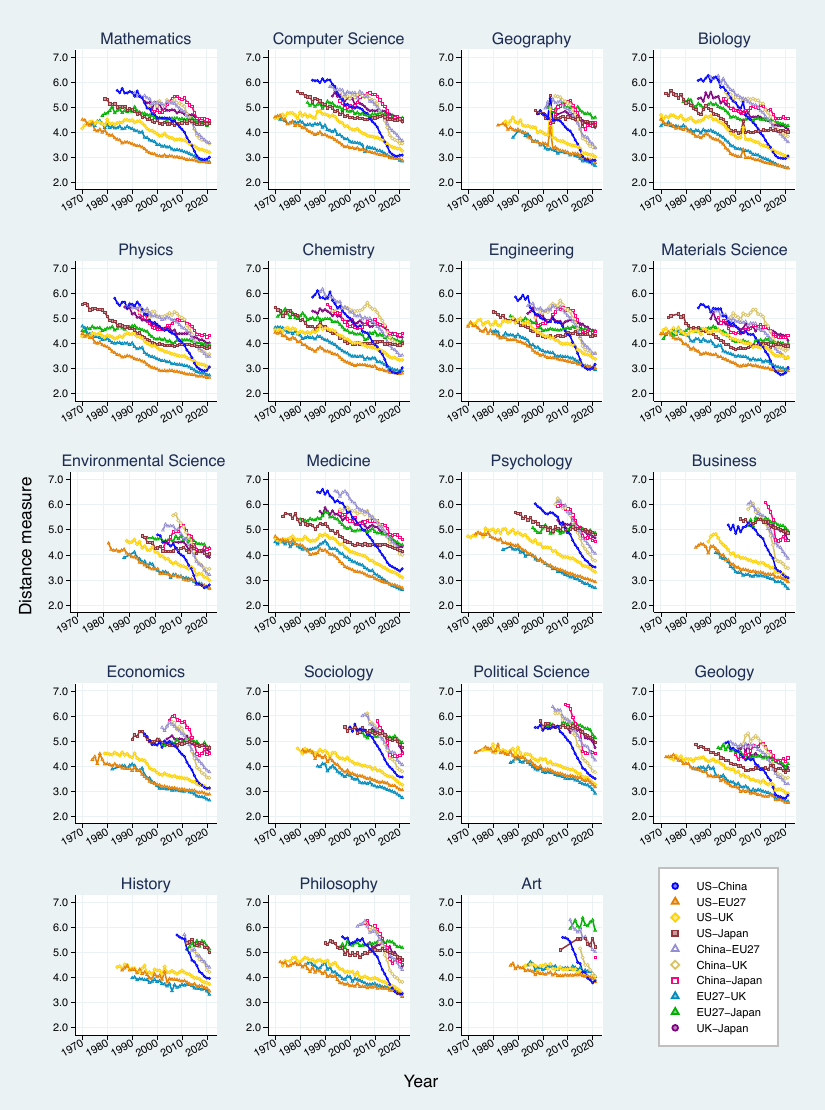}
\caption{\textbf{Change in the Collaboration Distance among the five parties over time.}
The changes in the distance from 1970 to 2021 between the five parties---the US, China, EU27, the UK and Japan---are shown for each of OpenAlex's 19 level-0 categories.}
\label{sfig:19F_5P}
\end{figure}

\begin{figure}[H]
\centering
\vspace{-0.5cm}
\includegraphics[align=c, scale=1.18, vmargin=0mm]{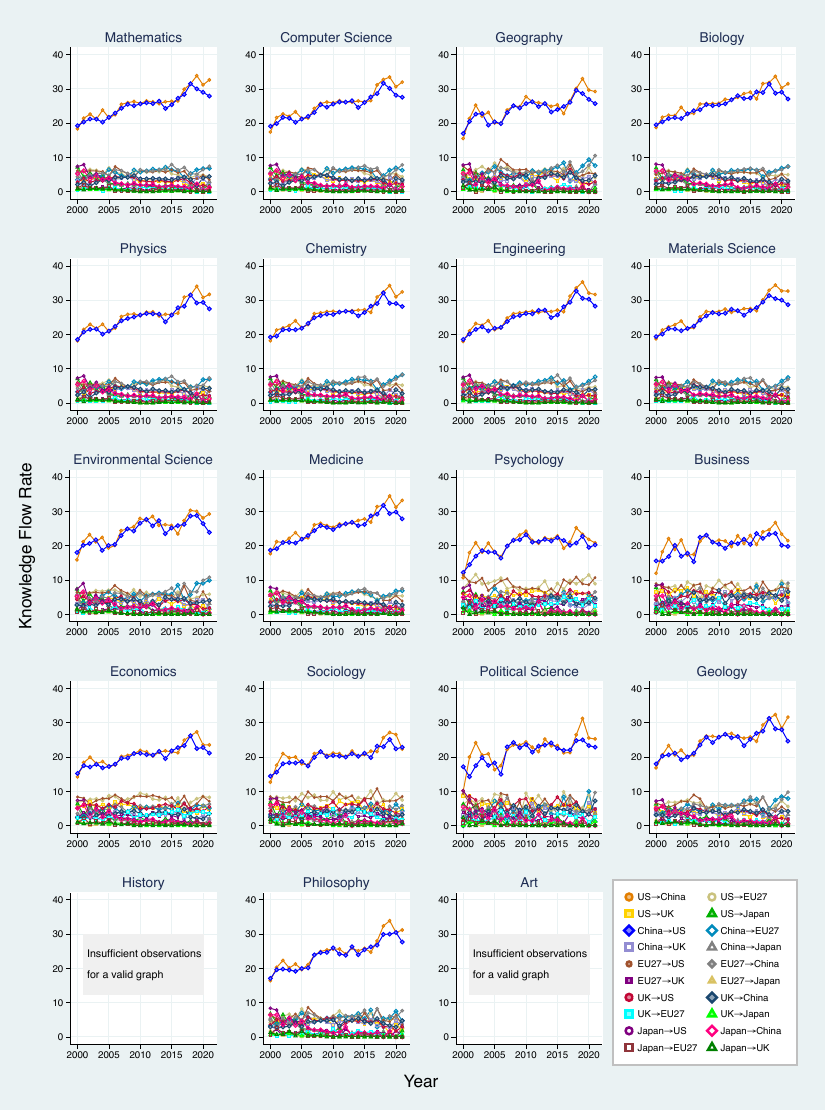}
\caption{\textbf{Change in the Knowledge Flow Rate among the five parties over time.}
The changes in the Knowledge Flow Rate from 2000 to 2021 between the five parties---the US, China, EU27, the UK and Japan---are shown for each of OpenAlex's 17 level-0 categories.}
\label{sfig:kflow_all}
\end{figure}

\begin{figure}[!t]
\centering
\vspace{-0.5cm}
\includegraphics[align=c, scale=0.7, vmargin=0mm]{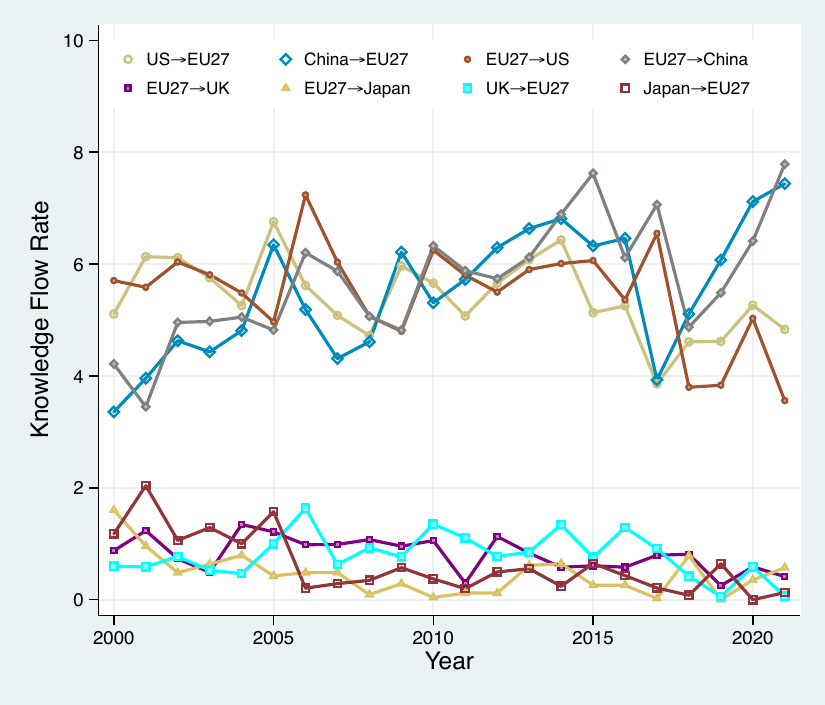}\\[0.5em]
\includegraphics[align=c, scale=0.7, vmargin=0mm]{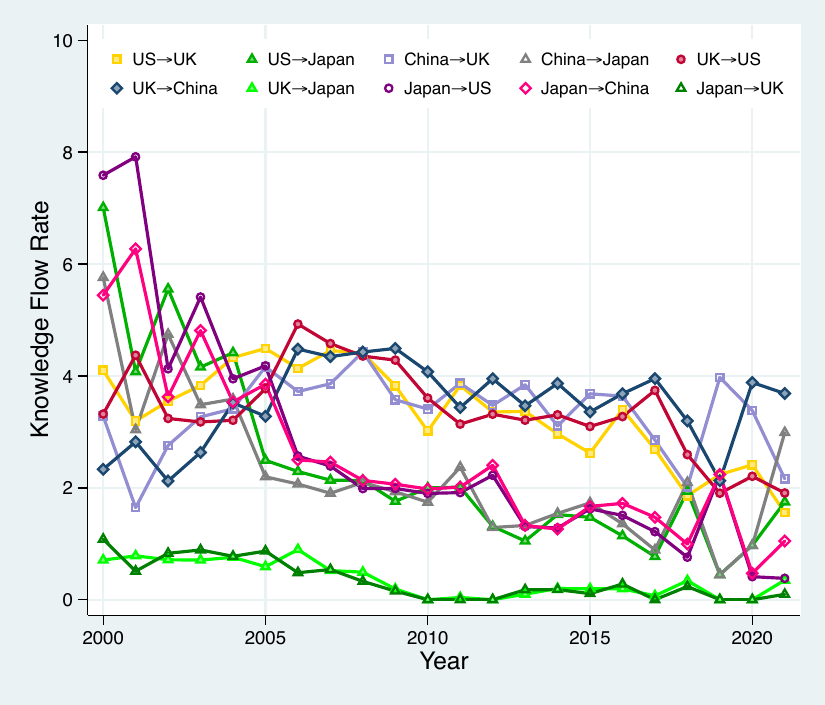}
\caption{\textbf{Change in the Knowledge Flow Rate among the scientific powerhouses in natural sciences.}
The outcomes pertaining to the US--China pair are illustrated in Fig.~\ref{fig:kflow_natall_USCN}, while the results for the remaining pairs are presented herein. 
}
\label{sfig:kflow_natall_Others}
\end{figure}
\quad\\
\vfill

\end{document}